\documentclass{sf2a-conf}
\usepackage{graphicx}
\usepackage{natbib}
\begin{document}

\TitreGlobal{SF2A 2008}

\title{NUV radii of the extrasolar planet HD~209458b}
\author{D\'esert, J.~M.$^1$ }
\author{Vidal-Madjar, A.$^1$ }
\author{Lecavelier des Etangs, A.$^1$}
\author{H\'ebrard, G.$^1$}
\author{Ferlet, R.}
\address{Institut d'Astrophysique de Paris, CNRS
(UMR~7095) Universit\'e Pierre \& Marie Curie; 98 bis, boulevard Arago 75014 Paris, France \\ email: {\tt desert@iap.fr} }

\runningtitle{NUV radii of HD~209458b}
\setcounter{page}{237}


\maketitle
\begin{abstract}

Extrasolar planetary transits are powerful tools to probe their
atmosphere and to extract key physical properties of planets,
like their mean densities, chemical compositions, or atmospheric
structures. Every 3.5 days, the transits of the gaseous planet
orbiting HD~209458 offer the opportunity to investigate the
spectral features of its atmosphere. We present here NUV
transmission spectroscopy of the transiting extrasolar planet
HD209458b using HST/ACS. We show the data analysis of the seven
HST orbits which were used to observe two transits of HD209458b.
Due to various remaining systematics, the radius of
the planet in the NUV could not be extracted with a high precison.
However, we derived a radius of $R_{P}=1.4~R_{jup} +/-0.08$ between 2800 and 3100\AA\ which is consistent with
previous measurements in the visible. 

\end{abstract}
%

\section{Introduction}

In the case of HD~209458b, absorptions of several percents for
H\,{\sc i} Lyman-$\alpha$, O\,{\sc i} and C\,{\sc ii} have been
measured in the hydrodynamically escaping upper atmosphere
(Vidal-Madjar 2003, 2004, 2008, D\'esert et al. 2004, Lecavelier
Des Etangs 2007, Ehrenreich et al. 2008) and a hot layer of
hydrogen have been detected (Ballester et al. 2007). More
recently, Rayleigh scattering by H$_2$ molecules has been
identified (Lecavelier et~al. 2008b) from sets of HST/STIS
observations (Charbonneau et al. 2002, Knutson et al. 2007, Sing
et~al. 2008a,b). From the same datasets, the possible presence of
TiO/VO as been studied (D\'esert et al. 2008).

\section{Observations}

The observing program was originally designed to observe
HD~209458b with HST/STIS. After STIS failure, it has been possible
to execute the program with HST/ACS during Cycle 13. The program
(GO10145) consists in $3+2$ visits, performed with the ACS/HRC and
the ACS/SBC. The results obtained using ACS/SBS with PR110L prism
spectroscopy around the Lyman-$\alpha$ line are presented in a
separated paper (Ehrenreich et la. 2008). The two other visits are
composed of $7$ orbits ($4+3$). Each orbit was plan to be  a
sequence of direct image filters and prism PR200L spectra at
various exposure times. The normal observing technique for all
ACS-HRC PR200L spectroscopy is to obtain a direct image of the
field followed by the dispersed prism image. This combination
allows the wavelength calibration of individual target spectra by
reference to the corresponding direct images. We obtain few
exposures of 0.1s, 3.6 s, 40 s, and 540 s during the two transit
observed. All the images were saturated.

\section{Analysis}

\subsection{Slitless spectroscopy}

HRC PR200L Slitless spectroscopy yields spectral 2D calibrated
images. The spectral image is composed of the stellar halo
surrounding the stellar center position, the red pile-up
resulting from the built up of photons on a few detector pixels
which appear when using a prism the saturation zone and the
stellar continuum between 150 and 400 nm. The spectral resolution
significantly diminishes toward the red. For bright objects, such
as HD~209458, this effect can lead to blooming of the HRC CCD from
filled wells; the overfilled pixels bleed in the detector Y
direction. The tilt of the prism causes a deviation of about 300
pixels between the position of the direct object and the region of
the dispersed spectrum on the CCD.

\subsection{Spectra extraction}
All the effect described previously make this extraction very
difficult. We used the flat-fielded spectral images to extract the
spectra. Since there is no slit in the ACS, the Point Spread
Function of the target modulates the spectral resolution. The so
called background is the sum of the effects of the real
background, the halo, and the diffraction spikes. Due to the quasi
central symmetry of the image, we considered that a good
evaluation of the background was to consider the opposite pixels
from the star. The current calibration of the wavelength solution
used in the aXe data reduction software assumes the use of these
apertures. At 3500 \AA, the dispersion drops to 105 \AA\ per pixel
Lyman-$\alpha$and is 563 \AA\ per pixel at 5000 \AA.

\subsection{Fitting the Transit Light Curve}

We parameterized the transit light curve with 4 variables: the
planet-star radius ratio $R_p / R_\star$, the stellar radius to
orbital radius ratio $a / R_\star$, the impact parameter $b$, and
the time of mid-transit $T_c0$. We used the transit routine
\texttt{OCCULTNL} developed by Mandel \& Agol (2002) where
limb-darkening corrections are taken into account. We then
performed a least-squares fit to our data over the whole parameter
space.

\section{Results and conclusion}

We extracted the NUV radius of HD~209458b at 2 bandpasses. We are
able to detect the stellar MgII doublet at ~ 2\,800 \AA. We found
a radius of $R_{P}=1.4~R_{jup} +/-0.08$ in agreement with the
radius of $1.3263~R_{jup}+/-0.0018$ found in the visible (Knutson
et al. 2007) obtain from STIS/HST observations.

Although ACS is very sensitive, the error bars derived here are
large since they include various systematics that affect the
determination of the radius of the planet. The telescope jitter,
dithering and intra-pixel sensitivity variations are responsible
for severe fluctuations observed in individual spectra. We
conclude that dithering should not be used for precision
photometry when levels of 0.01\% are needed. Finally, the
exposures were all saturated which introduce not linearity effects
and thus make the comparison between consecutive wavelength
difficult. The error bars we derived are to large to draw any firm
conclusion on the detection of absorbers potentially present in
the atmosphere and which could be observed in this wavelength
domain.

COS will reach the domain of 1200-3000 \AA. This domain include
the ACS one but with a much better resolution and sensitivity,
allowing the detection of supplementary absorber in this region.



\end{document}